\documentclass[a4paper,fleqn]{cas-dc}

\usepackage[square,sort,comma,numbers]{natbib}
\usepackage{pict2e}
\usepackage{amsmath}
\usepackage{lipsum}
\usepackage{nomencl}
\usepackage{comment}
\usepackage{lettrine}

\makenomenclature
\def\tsc#1{\csdef{#1}{\textsc{\lowercase{#1}}\xspace}}
\tsc{WGM}
\tsc{QE}
\tsc{EP}
\tsc{PMS}
\tsc{BEC}
\tsc{DE}
\setcitestyle{square}

\begin{document}
\let\WriteBookmarks\relax
\def\floatpagepagefraction{1}
\def\textpagefraction{.001}

\shorttitle{Local Cold Load Pick-up Estimation Using Customer Energy Consumption Measurements}

\shortauthors{S Bajic et~al.}

\title [mode = title]{Local Cold Load Pick-up Estimation Using Customer Energy Consumption Measurements}                      

\author[inst1,inst2]{Sanja Bajic}
\affiliation[inst1]{organization={Department of Electrical and Computer Engineering, McGill University},
            city={Montreal},
            postcode={H3A 0E9}, 
            state={QC},
            country={Canada}}
\author[inst1,inst2]{François Bouffard}
 \affiliation[inst2]{organization={Groupe d'études et de recherche en analyse des décisions (GERAD)},
             city={Montreal},
             postcode={H3T 2A7}, 
             state={QC},
             country={Canada}}
\author[inst1]{Hannah Michalska}

\author[inst1]{Géza Joós}



\nonumnote{}

\begin{abstract}
Thermostatically-controlled loads have a significant impact on electricity demand after service is restored following an outage, a phenomenon known as cold load pick-up (CLPU). Active management of CLPU is becoming an essential tool for distribution system operators who seek to defer network upgrades and speed up post-outage customer restoration. One key functionality needed for actively managing CLPU is its forecast at various scales. The widespread deployment of smart metering devices is also opening up new opportunities for data-driven load modeling and forecast. In this paper, we propose an approach for customer-side estimation of CLPU using time-stamped local load measurements. The proposed method uses Auto-Regressive Integrated Moving Average (ARIMA) modeling for short-term foregone energy consumption forecast during an outage. Forecasts are made on an hourly basis to estimate the energy to potentially recover after outages lasting up to several hours. Moreover, to account for changing customer behavior and weather, the model order is adjusted dynamically. Simulation results based on actual smart meter measurements are presented for 50 residential customers over the duration of one year. These results are validated using physical modeling of residential loads and are shown to match well the ARIMA-based forecasts. Additionally, accuracy and execution speed has been compared with other state-of-the-art approaches for time-series forecasting including  Long Short Term Memory Network (LSTM) and Holt-Winters Exponential Smoothing (HWES). ARIMA-based forecast is found to offer superior performance both in terms of accuracy and computation speed.
\end{abstract}



\begin{keywords}
cold load pick-up\sep demand forecasting\sep  distribution systems\sep power system restoration
\end{keywords}

\maketitle

\mbox{}
\nomenclature[01]{$a_{t}$}{Residual error at time $t$}
\nomenclature[02]{$B$}{Backshift operator}
\nomenclature[06]{$d$}{Order of differentiation}
\nomenclature[07]{$E_{C}$}{Net CLPU energy}
\nomenclature[08]{$E_{n}$}{Energy consumption in normal conditions - without an outage}
\nomenclature[11]{$E_{0}$}{Energy not served during the outage}
\nomenclature[13]{$n_{acf}$}{Number of significant lags in auto-correlation function}
\nomenclature[15]{$n_{pacf}$}{Number of significant lags in partial auto-correlation function}
\nomenclature[17]{$p$}{Auto-regressive model order}
\nomenclature[17]{$p_{max}$}{Maximum value of the order of auto-regression}
\nomenclature[17]{$p_{limit}$}{Limiting value of the model order of auto-regression}
\nomenclature[20]{$P_{CLPU}$}{Maximum CLPU power consumption}
\nomenclature[25]{$P_{no}$}{Power consumption in normal operating conditions}
\nomenclature[55]{$P_{peak}$}{Daily peak value}
\nomenclature[70]{$q$}{Moving average model order}
\nomenclature[70]{$q_{max}$}{Maximum value of moving average order}
\nomenclature[70]{$q_{limit}$}{Limiting value of the moving average model order}
\nomenclature[75]{$t_{C}$}{CLPU duration}
\nomenclature[76]{$t_{0}$}{Outage start time}
\nomenclature[77]{$\phi_{i}$}{Coefficient of $t$-$i^{th}$ auto-regressive term in ARIMA model}
\nomenclature[78]{$X_{t}$}{Measured time series values at time $t$}
\nomenclature[79]{$W_{t}$}{Differenced measurement series at time $t$}
\nomenclature[90]{$\theta_{i}$}{Coefficient of $t$-$i^{th}$ moving average variable term}
\nomenclature[91]{$\tau_{1}$}{Outage end time}
\nomenclature[92]{$\tau_{2}$}{CLPU end time}
\nomenclature[93]{$\varphi_{i}$}{Coefficient of $t$-$i^{th}$ auto-regressive term in peak power estimation}
\nomenclature[89]{$\beta_{i}$}{Coefficient of $t$-$i^{th}$ lagged control variable term}

\printnomenclature

\section{Introduction}
\label{sec:sample1}
\lettrine{U}{tilities}, whose customers have large thermostatically controlled loads (TCL), face significant challenges due to sharp load increases when service is restored following extended outages. This is especially important in regions where and when winter temperatures are low (e.g. $-40 ^ { \circ } \mathrm { C }$), as in Quebec, Canada \cite{dzeletovicreduction}.  Similar comments apply during heatwaves for power systems with significant air conditioning loading. At the same time, it is not unreasonable to expect that a deeper penetration of electric vehicle charging loads will exacerbate these challenges \cite{visakh2023}.

The sudden and significant increase in the load after service restoration, referred to as \textit{cold load pick-up} (CLPU), is principally attributed to loss of load diversity. CLPU is habitually accounted for in planning and operation procedures; in fact, it is one of the top determining factors of high-to-medium voltage transformer sizing and their overload capability \cite{Lefebvre2002}, \cite{dzeletovic2019}. Utilities are actively working on the improvement of strategies to reduce CLPU demand and in that way reduce the peak loading and thus potentially postpone investments.

The motivation behind this work is found in solving several practical problems that more and more utilities are facing:
\begin{enumerate}
    \item Utilities use outdated CLPU values derived by rules of thumb, where the peak power is multiplied by a constant depending on the customer type. As customer behavior is complex, this value can be either an overestimate of the real demand, in which case utility is not using the grid efficiently, or an underestimate of the real CLPU demand. The latter is much more problematic as it can cause damage to the equipment and unnecessary relay protection switching and thus may leave many customers without electricity for even longer periods.
    \item There are limited numbers of SCADA measurements on feeders as the installation of feeder measurements can be costly; as a result utilities seek to use all available data sources to predict network behavior. Additionally, the deadband on feeder measurements can be wide, and small changes go undetected, which can pose a challenge in the case when the cold load pick-up effect impacts low-voltage equipment and medium to low-voltage transformers. Therefore, estimating CLPU at the end-user level is necessary due to the lack of measurements and accurate physical models of end-customer behavior.
     \item In the coming years, utilities are expected to increasingly decentralize decision-making processes to reduce their reliance on centralized control and facilitate the integration of distributed energy resources.
     
     In the same vein, to effectively implement restoration algorithms with explicit CLPU management as proposed, for example, in \cite{dzeletovic2019}, it is essential to have a handle on how much energy single customers or small groups of customers (e.g., all customers connected to the same medium- to low-voltage transformer) need to recoup once service is restored. With close to real-time estimates of ``missing'' energy, distribution system operators can modulate their restoration strategy accordingly to smooth out feeder overloads while potentially providing preferential restoration to customers needing it the most \cite{dzeletovic2019}.
\end{enumerate}

To address the challenges listed above, this paper proposes a methodology for estimating CLPU of a single residential dwelling\footnote{This could also be done for a few dwellings sharing a distribution transformer.} with significant amounts of TCLs based on smart meter measurements (timestamped demand records). Without loss of generality here, the main focus is to achieve optimal estimation performance for customers who use electrical heating. During low-temperature periods, the CLPU of such customers poses an additional challenge for the distribution utility in both operations and planning. The proposed approach is based on the estimation of future \emph{energy} consumption for the duration of the potential outage using time-series forecast by fitting an Auto-Regressive Integrated Moving Average (ARIMA) model. We will argue that in spite of their simplicity, ARIMA models are well-suited for this application due to their low computation needs, a feature essential for local estimation. It is also important to highlight that the ARIMA model here is meant to predict \emph{energy}, not \emph{power}. This is a positive feature of this approach because energy consumed over time is a much smoother random process than power over time. Moreover, ARIMA model parameters are updated hourly to adapt to changing conditions (e.g., changing outside temperatures) that may affect demand and the value of CLPU. Model order selection is performed for the most recent data sample based on the analysis of data stationarity, values of the auto-correlation function (ACF) and partial auto-correlation function (PACF). 

The remainder of this paper is organized as follows. In Section 2, we present a literature review of CLPU phenomena and short-term energy consumption forecast. Next, we present the architecture of local CLPU estimation in Section 3. In Section 4, we develop the mathematical formulation for the ARIMA model and model selection. Section 5 looks at evaluating CLPU ARIMA model selection efficacy on actual smart meter data and validating CLPU duration assessment against simulated cases based on explicit load models implemented in GridLab-D, an industry-grade distribution system simulation software. Additionally, in this section, we compare ARIMA models with two state-of-art methods for time-series prediction, namely Long Short Term Memory Networks (LSTM) and Holt-Winters Exponential Smoothing (HWES). The conclusions of the work follow in Section 6.

\section{Literature Review}
\label{sec:sample1}
Previous work on the subject of CLPU has focused on physically-based load modelling and post-restoration load response. Physically-based CLPU models of heating and cooling loads were first developed in \cite{ihara1981}. In \cite{Schneider2016}, a multi-state physical model is employed to evaluate the magnitude and duration of CLPU. Physically-based models for computing electric loads on residential feeders are presented in \cite{Lefebvre2002}. A feeder CLPU curve for an aggregated load consisting of single-family homes with electrical space and water heating is computed as a piecewise linear function of outdoor temperature by the authors of \cite{Aubin1991}. An aggregate model for a heterogeneous population of TCLs is derived in \cite{Zhang2012} to capture accurately the collective load behaviour under demand response. All of the above models were developed for aggregated loads, and thus cannot be readily applied for single residential CLPU estimation. In \cite{Penaranda2022}, a state space model is used to determine CLPU demand of an end-use load. Despite the high accuracy of physically-based load modeling, they are inherently difficult to build and maintain because of their significant and changing model calibration requirements.

Hence, measurement-based approaches show better potential for CLPU modeling. The CLPU model of aggregate residential loads developed in \cite{agneholm2000} is based on field measurements and a load model. In that article, the authors introduce the CLPU factor that represents the ratio between the rated power and the expected power during the restoration and quantifies the energy to be recovered which depends on the duration of an outage. A general model for determining the maximum restorable load was developed in \cite{Qu2012}. Here, CLPU is estimated using only \textit{practical guidelines}---rules defined based on the practical experiences for accounting for CLPU peak and duration in the service restoration procedure. Practical guidelines for CLPU estimation have routinely been applied in many utilities, as they proved to be sufficiently robust and effective provided that guideline assumptions were satisfied. The drawback of this approach, however, is that it does not apply to the forecast of a single load CLPU peak and duration. The same drawback applies to \cite{meng2021}, where authors recognize the necessity of a time-dependent CLPU model for the aggregated load. The aforementioned represents a limitation in novel approaches for CLPU mitigation strategies, as the assumptions are built based on the aggregated load response. In \cite{chuan2022}, authors mention this limitation and approach it by applying a feeder-level exponential decay CLPU model on a single dwelling. Such simplification can introduce underestimation of a single load CLPU considering that the feeder-level CLPU model has a significantly larger load diversity factor.

Recently, there has been an increase in deployment, and consequently, the utilization of timestamped customer load measurements. The authors of \cite{bu2019} developed a data-driven framework for the estimation of CLPU response of aggregated load. The timestamped customer load measurements are used to enhance the results of a least square support vector machine. In \cite{dzeletovic2020}, timestamped customer load measurements along with outdoor temperature measurements are used to estimate energy to be recovered after an outage. Compared to the temperature-based linear regression approaches, auto-regression has shown better performance in terms of mean squared error (MSE). This has been identified in \cite{Menazzi2022}, where the pre-fault smart meter measurements serve as an input to a machine learning integrated CLPU model to obtain short-term load forecasts. The disadvantage of the model proposed here is that it is based on several assumptions that are very difficult to fulfill in practice; these include knowing a priori the CLPU parameters of the customer, the ratio of TCL loading to the total load at the initiation of the outage, and that the customer keeps thermostat settings constant.

One of the determining factors of CLPU size and duration is the amount of energy not served to a customer during an outage. This energy can be predicted using energy consumption measurements. The massive smart meter deployment over the past decade has provided the industry with large amounts of data which are highly granular, both temporally and spatially \cite{hong2016probabilistic}. A comparison of the four most common approaches was performed in \cite{bonetto2017machine}: Auto-Regressive Moving Average (ARMA) models, Support Vector Machine (SVM) models, and two approaches based on Artificial Neural Networks (ANN): nonlinear auto-regressive recurrent neural networks, and long-term short-memory networks. Here, ARMA was outperformed by the other approaches. However, this conclusion is questionable since an ARMA model assumes stationary time series data, which is generally not the case for timestamped household load measurements. An ARIMA model would be a more suitable choice since it can be applied to non-stationary time-series data.

A detailed forecast model based on four approaches (multiple linear regression (MLR), regression trees, support vector regression (SVR) and ANN) is proposed in \cite{lusis2017short}. Here, a large number of features is introduced, such as temperature, humidity, precipitation rate, wind chill, average load values in several past intervals, to generate forecasts for every hour. Inasmuch this approach is probably superior in accuracy in comparison to competing approaches, its additional complexity can make one doubt of its applicability in an industrial setting.

In \cite{haben2019short} short-term load forecasts using smart meters have been implemented and compared for state-of-art forecasting methods for both point and probabilistic approaches. The results have shown that accurate forecasts can be obtained by relatively simple methods and that the best methods were those based on auto-regression and Holt-Winters-Taylor exponential smoothing. In this paper, we apply these methods in the context of CLPU estimation, i.e. \emph{missing energy} not \emph{missing power} which is a different problem in comparison to traditional load (power) forecasting. In \cite{nasqash2022}, authors have done an extensive literature review on load-foresting techniques for power systems where all the state-of-the-art techniques were analyzed, along with their advantages and disadvantages. The authors claim that the forecasting of time series can be successfully performed by statistical methods. Additionally, they conclude that ARIMA modeling is best suited for representing dynamic time series. Therefore, ARIMA models have several advantages that make them more suitable for short-term energy consumption forecast in comparison to alternative approaches:
\begin{enumerate}
    \item As a statistical forecasting approach, ARIMA has shown better properties in terms of accuracy and computational performance than Recurrent Neural Network (RNN) approaches \cite{Nichiforov2017}, \cite{makridakis2018statistical}. ARIMA is designed for time series data while RNN-based models are designed for sequence data which makes them harder to build out-of-the-box \cite{Siami2018}.
    \item ARIMA has shown good forecasting results in univariate modelling \cite{arora2013}. Univariate modelling of electrical energy consumption is more convenient for local and independent execution since it does not require additional data apart from timestamped load measurements available locally. Moreover, when doing local energy forecasting, computing power has to remain limited, especially if forecast has to be done in a grid edge device like a meter.
    \item Finally, ARIMA modeling does not assume knowledge of any underlying model or relationships, which makes it a good choice for predicting models with uncertainty \cite{adebiyi2014comparison}.  
\end{enumerate}

\section{Local CLPU Assessment Architecture}
In this section, we present the architecture of the proposed framework for local CLPU estimation. The method is designed for local execution at the service delivery point, fed by the most recent local smart meter data. The outputs of the algorithm are forecasts of the CLPU peak and the duration of CLPU events following a range of outage durations.

\subsection{Single Load CLPU}
\begin{figure}[!tbp]
  \centering
  \begin{minipage}[b]{0.45\textwidth}
    \includegraphics[width=\textwidth]{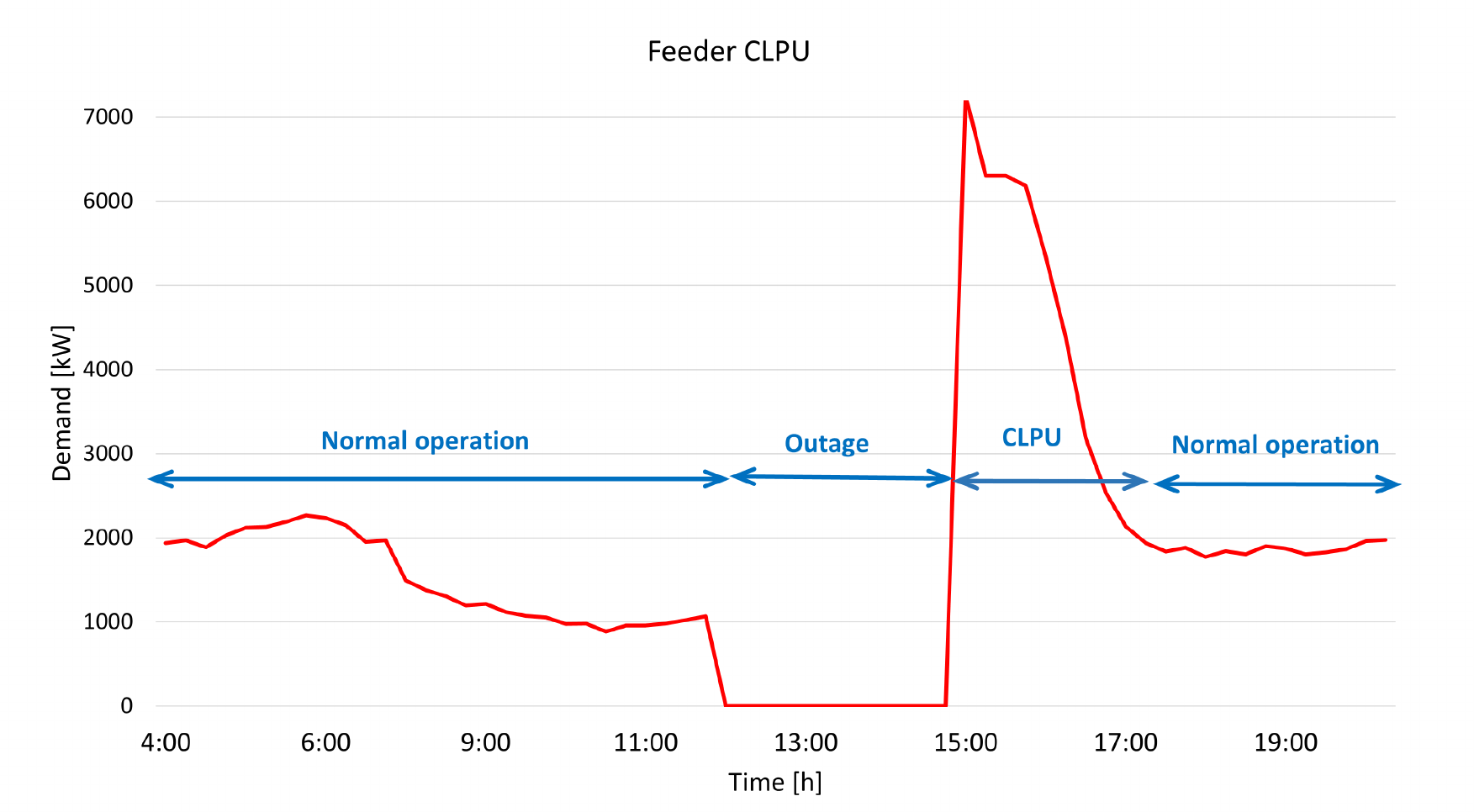}
    \caption{Feeder CLPU.}
    \label{fig:feed}
  \end{minipage}
  \hfill
  \begin{minipage}[b]{0.45\textwidth}
    \includegraphics[width=\textwidth]{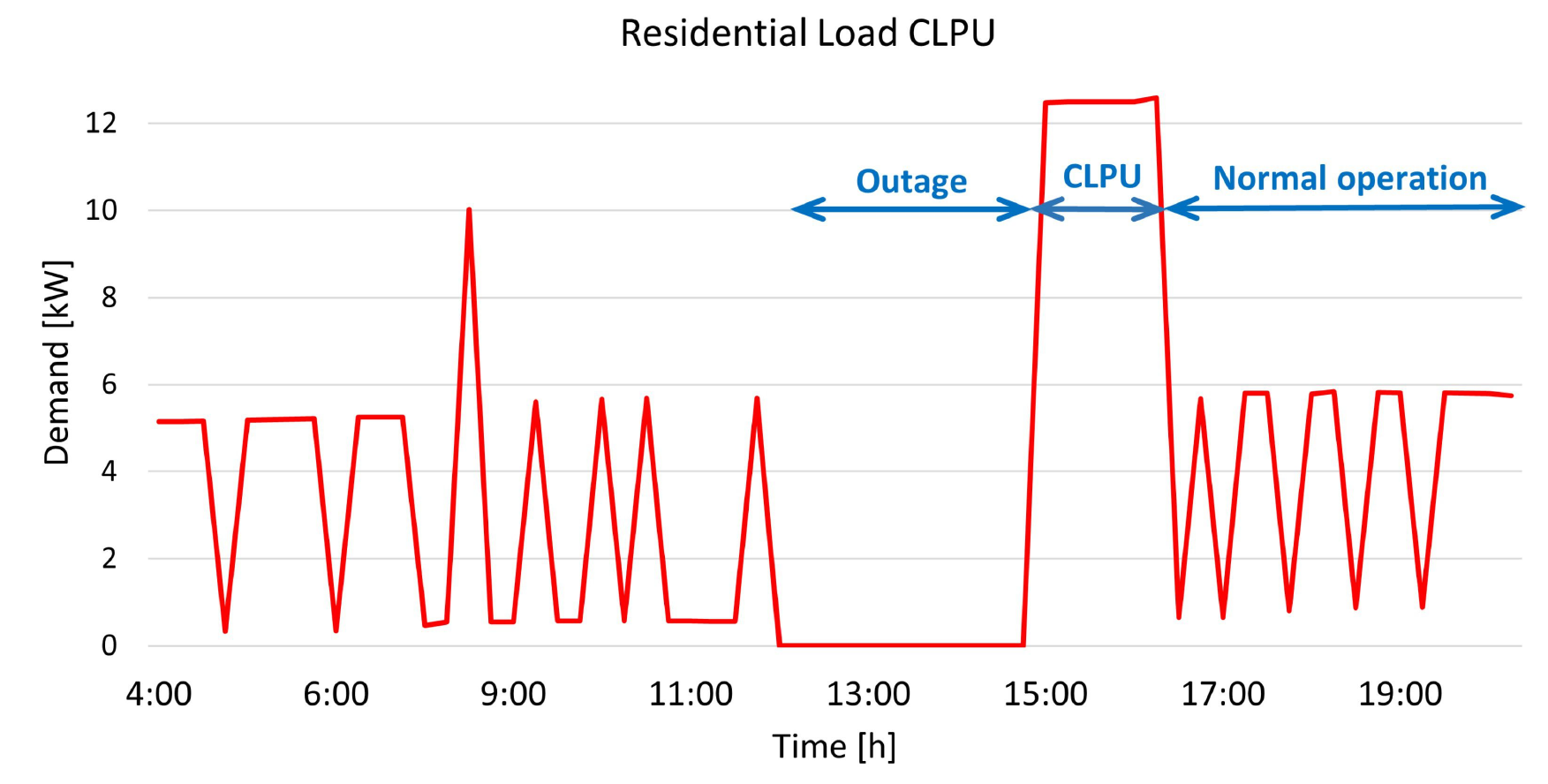}
    \caption{Single dwelling CLPU.}
    \label{fig:load}
  \end{minipage}
\end{figure}

Due to the significant difference between a CLPU curve at the feeder level and at the single dwelling level, we should approach differently the estimation of CLPU for these two cases.  Fig. \ref{fig:feed} shows a typical feeder load during a CLPU event. On the other hand, Fig. \ref{fig:load} illustrates CLPU of a single residential dwelling during the same outage as in Fig.~\ref{fig:feed}. By inspection, we can see that feeder-level CLPU displays exponential decay, while the single customer CLPU is discrete in nature (i.e., all or nothing) due to its low number of appliances.

For a single dwelling, CLPU characterization boils down to two decoupled problems:
\begin{enumerate}
    \item Estimation of the peak demand during CLPU,
    \item Estimation of energy not served during the outage, which is proportional to the time required to regain normal operation.

\end{enumerate}

\begin{figure}
\centering
\includegraphics [width=0.45\textwidth] {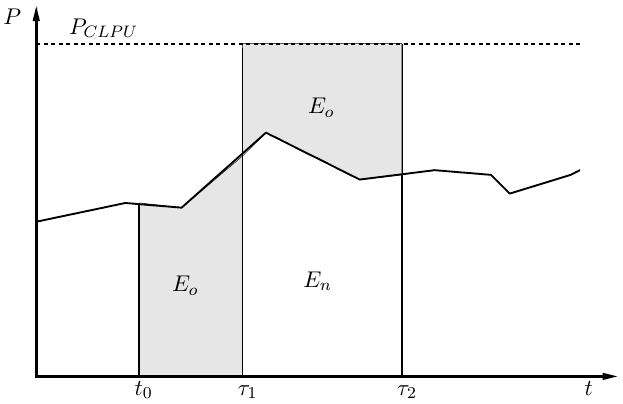}
\caption{CLPU estimation using the estimated value of energy not served during the outage.}
\label{fig:etr}
\end{figure}

Due to the discrete nature of residential CLPU, the duration of the CLPU peak demand varies with the energy not delivered during the outage. This assertion is based on the fact that during an outage the temperature of a dwelling will drop (rise) due to the lack of space heating (cooling). Ambient temperatures are returned to their set points only once the energy that would have been spent during the outage to heat (cool) a dwelling is recovered. The peak power, on the other hand, is limited by the maximum power of the dwelling's major TCLs and, possibly, its electric vehicle charger. Consequently, in this paper, we address the CLPU forecast for a single customer by first estimating the energy not delivered during the outage (lasting from $t_0$ to $\tau_1$) and the peak power, $P_{CLPU}$, which is the rate of energy use in the post-outage period (lasting from $\tau_1$ to $\tau_2$), as illustrated in Fig. \ref{fig:etr}. In this figure, the curve represents the energy consumption in normal operating conditions when there is no outage.

If an outage starts at $t_{0}$, the CLPU duration $t_{C}=\tau_{2}-\tau_{1}$ depends on the outage duration time, $\tau_1 - t_0$, due to the need to recover energy lost during the outage:
\begin{equation}
    E_{C}(t_0, \tau_1, \tau_2) = E_{o}(t_0, \tau_1) + E_{n}(\tau_1, \tau_2)
\end{equation}
where $E_{C}$ is the net CLPU energy, $E_{o}$ is the unsupplied energy during the outage and $E_{n}$ represents the ``normal'' energy consumption (i.e., as if there had been no outage) during the CLPU period. Using the anticipated energy consumption in normal operating conditions, $E_{o}$, and the amount of energy to recover $ E_{C} $, we can estimate the time needed for return to normal operation, by estimating the time to consume the $E_{C}$ at the CLPU power $P_{CLPU}$.

\subsection{CLPU Peak}
The maximum power consumption of a single load is limited by the contract with the utility and the rating of household protection equipment. Daily peak power values depend on the sum of the ratings of the appliances likely to be used simultaneously. The dominant part of the load in the low- (or high-) temperature periods consists of electrical heating (cooling) units.\footnote{For the remaining parts of the paper, we will only refer to heating loads to be the leading cause of CLPU in cold weather. The converse is applicable during heat waves to cooling loads.} Both the temperature dependency of the heating load and household appliances' power consumption influence the daily maximum load power. Hence, the power in the post-outage periods of a single load, due to the load power limitations, can be approximated by the daily peak power. Furthermore, we can calculate maximum daily power by applying an autoregressive model to the historical daily maximum power measurements
\begin{align}
P_{CLPU}(t) &= \hat{P}_{CLPU}(t) + a(t) \nonumber \\ &= \sum_{i=1}^{N} \varphi_{i} P_{peak}(t-i)+a(t) \label{eq:peak}
\end{align}
where peak demand in the CLPU conditions $P_{CLPU}$ at time $t$ is estimated by $\hat{P}_{CLPU}$ using an autoregressive model of peak power values $P_{peak}$ of the previous days, with coefficients $\varphi_{i}$ and residual error $a(t)$.

\subsection{CLPU Duration}
As mentioned above, we define the CLPU duration of a single customer $t_C$ as the time for which the customer has continuous energy consumption at a high power value, as shown in Fig. \ref{fig:load}. The estimate of the CLPU duration is calculated using the estimated CLPU peak power $\hat{P}_{CLPU}(t_0)$ valid at the initiation of the outage $t_0$ and the forecast of the unsupplied energy $\hat{E}_{o}(t_0, \tau_1)$, that is
\begin{equation}
   \hat{t}_C (t_0, \tau_1) \approx \frac{\hat{E}_o(t_0, \tau_1)}{\hat{P}_{CLPU}(t_0)} \label{eq:duration}
\end{equation}
where now we express $\hat{t}_C$ to be a function of both the outage initiation time $t_0$ and its time to restoration $\tau_1$. We note that this estimate is computed a discrete number of times every hour for a number of outage duration times $\tau_1 - t_0$. In the following section, we will address how the estimate $\hat{E}_o(t_0, \tau_1)$ is determined.

\subsection{CLPU Energy}

As mentioned in the introduction, the challenge with energy time series forecast is the non-stationarity of this random process. This is why ARIMA models are well-suited for this forecast task.

The random process $E_o(t_0, \tau_1)$ is non-stationary by nature: energy consumption at the individual household is periodic because of daily energy use cycles of its occupants. This is reinforced by how diurnal weather patterns influence space heating (or cooling) needs of a building. Therefore, differencing of past energy use is needed to prior to the determination of an ARIMA model for $E_o(t_0, \tau_1)$. For this, we define differenced energy record at time $t$
\begin{equation}\label{ARIMA:1}
    \tilde{E}_o(t) = E_o(t) - E_o(t-d)
\end{equation}
where $d$ is the order of the differencing applied to the energy use records of the customer.

From the differenced energy consumption time series, one can infer the following ARIMA model for the energy consumed in the next energy measurement time interval (i.e., the time between the logging of two energy consumption values, which ranges typically between 15 and 30 minutes in modern advanced metering infrastructures)
\begin{equation}\label{ARIMA:2}
    E_o(t) = b(t) + E_o(t-d) + \sum_{i=1}^p \phi_i \tilde{E}_o(t-i) + \sum_{i=1}^q \theta_i b(t-i)
\end{equation}
where
\begin{equation}\label{ARIMA:3}
    b(t) = E_o(t) - \hat{E}_o(t)
\end{equation}
is the error between the true energy consumption at $t$ and its forecast $\hat{E}_o(t)$
\begin{equation}\label{ARIMA:4}
    \hat{E}_o(t) = E_o(t-d) + \sum_{i=1}^p \phi_i \tilde{E}_o(t-i) + \sum_{i=1}^q \theta_i b(t-i)
\end{equation}
The parameters $\phi_1, \phi_2, \ldots, \phi_p$ ($p \geq 1)$ and $\theta_1, \theta_2, \ldots, \theta_q$ ($q \geq 1$) are the ARIMA model parameters to be fitted based on historical energy consumption records. We shall discuss the determination of the optimal values of the order parameters $d$, $p$ and $q$ in the next section.

Based on \eqref{ARIMA:1}--\eqref{ARIMA:4}, we posit that energy not supplied during an outage starting at time $t = t_0$ and lasting until $t = t_0 + \tau_1 = t_0 + r \delta$, where $r \geq 0$ is an integer, and $\delta > 0$ is the time resolution of the CLPU energy calculation
\begin{align}\label{ARIMA:5}
    E_o(t_0,\tau_1) &= \sum_{j=0}^{r} \Big [ b(t_0+j) + E_o(t_0-d+j) \nonumber \\
    + & \sum_{i=1}^p \phi_i \tilde{E}_o(t_0-i+j) + \sum_{i=1}^q \theta_i b(t_0-i+j) \Big ]
\end{align}
This entails that the predictor of energy to be recovered if an outage starts at $t_0$ and ends at $\tau_1 \geq t_0$\footnote{In practice, $t_0$ and $\tau_1$ are discrete-valued. For instance, the time $t_0$ would typically correspond to the last time an energy measurement was logged by the meter prior to the onset of an outage. Similarly, as mentioned previously $\tau_1 = t_0 + r\delta$, where $\delta$ would typically be in the range of 15 to 30 minutes depending on the metering resolution in place. Moreover, it is also reasonable that the value of $r$ cannot grow too large, as one should expect the quality of the CLPU energy forecast to decrease as the forecast horizon increases.}

\begin{align}\label{ARIMA:6}
    \hat{E}_o(t_0,\tau_1) = \sum_{j=0}^{r}  \Big [E_o(t_0-& d+j) + \sum_{i=1}^p \phi_i \tilde{E}_o(t_0-i+j)  \nonumber \\
   & + \sum_{i=1}^q \theta_i b(t_0-i+j) \Big ]
\end{align}

We note that this forecast model can also be extended to include lagged ambient temperature values as shown in Appendix~\ref{appendix:arimax}.

The values of $\phi_{1},\phi_{2}, \ldots, \phi_{p}$ and $\theta_{0}, \theta_{1}, \ldots, \theta_{q}$ represent the coefficients of the model that are estimated from the logged energy consumption. The model is fully specified once the $p + q+ 1$ parameters have been determined from logged consumption data. Here, $d$ is the number of differences needed to render past consumption data stationary, $p$ is the number of auto-regressive terms, and $q$ is the number of lagged forecast errors.

\subsection{ARIMA Model Selection}
As mentioned previously, energy consumption time series are notoriously non-stationary---one can think of how the mean and the variance of the energy consumed varies from season to season. ARIMA models are tailored to address such lack of stationarity in the mean sense when compared to autoregressive (AR) and autoregressive moving-average (ARMA) models, which cannot since they can only model purely stationary processes.

One of the major challenges when using ARIMA models is to select an adequate model order, the value of three parameters: the number of auto-regressive terms ($p$), the number of moving average terms ($q$), and the order of differentiation of the time series ($d$).

The first parameter to be set is the differencing order, $d$. Intuitively energy consumption time series are periodic and do not display constant means over periods of time of a few hours. Therefore, they are non-stationary. Through the application of standard unit root tests like the Augmented Dickey-Fuller (ADF) test, it is possible to establish stationarity of a time series or lack thereof  \cite{armstrong2001principles}. For processes displaying time-variable means, it is common to apply differencing to remove its seasonal/periodic component. The order of differencing, $d$, sets the number of times the data has had earlier values subtracted from them. The ADF test is applied successively as $d$ is increased until the time series displays adequate stationarity characteristics.

The next step is to identify the values of $p$ and $q$ that would return the \emph{best} forecasting performance. By \emph{best} performance here, we entail in the mean-sense stationary based on the estimated time series autocorrelation function (ACF) and the partial autocorrelation function (PACF), defined in Appendix~\ref{appendix:acfpacf}. The task here is to carry out a search over the discrete combinations of $p$ and $q$ resulting in minimal forecasting error. For a given AR process order $p$, analysis of PACF is performed, while for the MA process order $q$, the ACF is calculated \cite{AileenNielsen2019}. 
Based on the number of significant lags in ACF and PACF, optimal values of lags for AR and MA processes are set.

\begin{figure}
\centering
\includegraphics [width=0.40\textwidth] {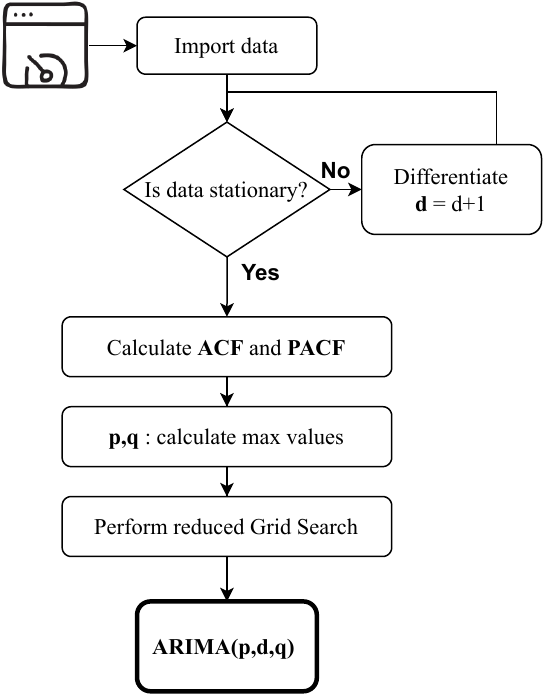}
\caption{ARIMA model order selection algorithm.}
\label{fig:ord}
\end{figure}

The simplest approach to determining optimal values for the model order is to perform a full grid search over all possible $p$ and $q$ values. The most suitable model order is selected based on some of the standard criteria, such as mean squared error (MSE), root mean squared error (RMSE), Akaike information criterion (AIC) and others \cite{AileenNielsen2019}. However, this is a computationally expensive process if there is significant uncertainty with respect to the required model orders. To overcome this challenge, a reduced grid search is developed in this paper. The reduced grid search is an approach that aims to reduce the search space of the ARIMA model order using the statistical properties of the historical data such as ACF, PACF and unit root test results. This is done by tracking the trends seen in the MSE as $p$ and $q$ are increased individually. As long as MSE drops singifcantly with increasing order, the maximal allowable model orders are upped. Once the decreases in MSE become insignificant with an order increase, maximum values for both $p$ and $q$ are established. Of course, this assumes that the optimal model order when considering $p$ and $q$ jointly is expected to be less than or equal to the maximum model orders for the AR and MA portions of the model when considered individually. Fig. \ref{fig:ord} summarizes the algorithm for model order selection.

\section{Implementation}

\begin{figure}
\centering
\includegraphics [width=0.45\textwidth] {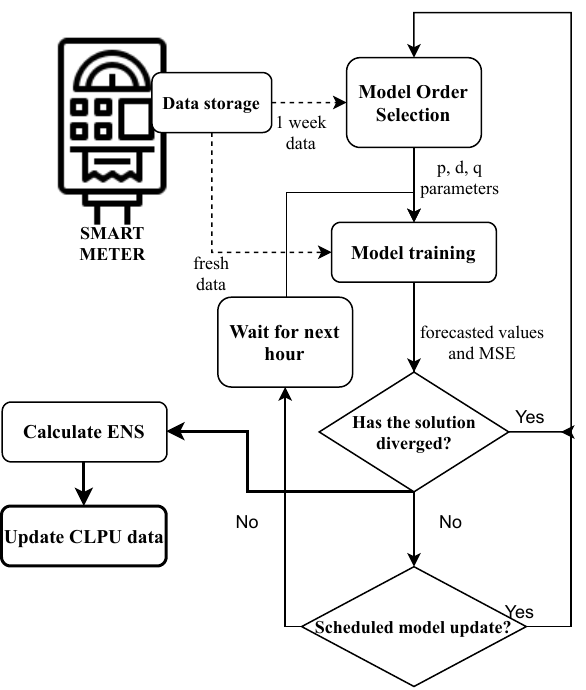}
\caption{CLPU assessment algorithm overview.}
\label{fig:arch}
\end{figure}

The proposed solution architecture is shown in Fig. \ref{fig:arch}. The CLPU forecasting system architecture consists of a smart meter with an additional calculation module that performs CLPU peak and duration estimation bases on the value of Energy Not Supplied (ENS). This module is available on-demand to the distribution system operator prior to outages, planned or not. The model order selection algorithm from which the ARIMA model is obtained, i.e. the values of parameters $p$, $d$ and $q$, is performed using fresh time series data on a periodic basis. The model order selection procedure is re-triggered on a weekly basis or in the case of the divergence of maximum likelihood estimation (MLE) optimization algorithm used for the estimation of ARIMA model parameters.

Using the same data, models with the specified orders is trained to obtain the parameters used for the forecasting of the CLPU peak ($\varphi_i, i = 1, \ldots, N$) and forgone energy/CLPU duration ($\phi_i, i = 1, \ldots, p$ and $\theta_i, i = 1, \ldots, q$) following an outage.  The calculated CLPU peak and duration models thus provide necessary information to facilitate strategic restoration actions when feeder capacity is limited and/or to favor the restoration of customers needing electricity the most (i.e., those with the largest CLPU energy) over those who can wait longer. Therefore, at each time step $t_0$, the trained models calculate i. the CLPU peak as per \eqref{eq:peak}, ii. the CLPU energy as per \eqref{ARIMA:4}, and iii. the CLPU duration as per \eqref{eq:duration} for a set of discrete outage durations $\tau_1 - t_0 = r\delta$, where $\delta$ would typically be 15 or 30 minutes and $r \geq 1$.

\section{Results and discussion}
In this section, we illustrate how CLPU characteristics of individual dwellings are determined from local smart meter data. Here, we assess CLPU forecast abilities for outages up to 12 hours in one-hour time steps.

\subsection{Data Set Description}
In this paper, actual time-stamped dwelling-level measurements are used for verification of the proposed approach. The measured load data for residential homes were obtained from the UMassTraceRepository \cite{Barker_smart}. They are part of the Smart* Home Data Set which contains data for 114 single-family apartments for the period from 2014 to 2016 measured in 15-minute time intervals. Specifically, we have randomly selected 50 apartments from the `Apartment' dataset.

\subsection{System configuration}
All calculations were implemented in Python, using \cite{seabold2010statsmodels}, and ran on a computer with the following configuration: Intel Core i7 8700 CPU 3.19 GHz processor, 16.0~GB of RAM. 

\subsection{ARIMA Model Selection -- Performance Results}
\begin{figure}
\centering
\includegraphics [width=0.5\textwidth] {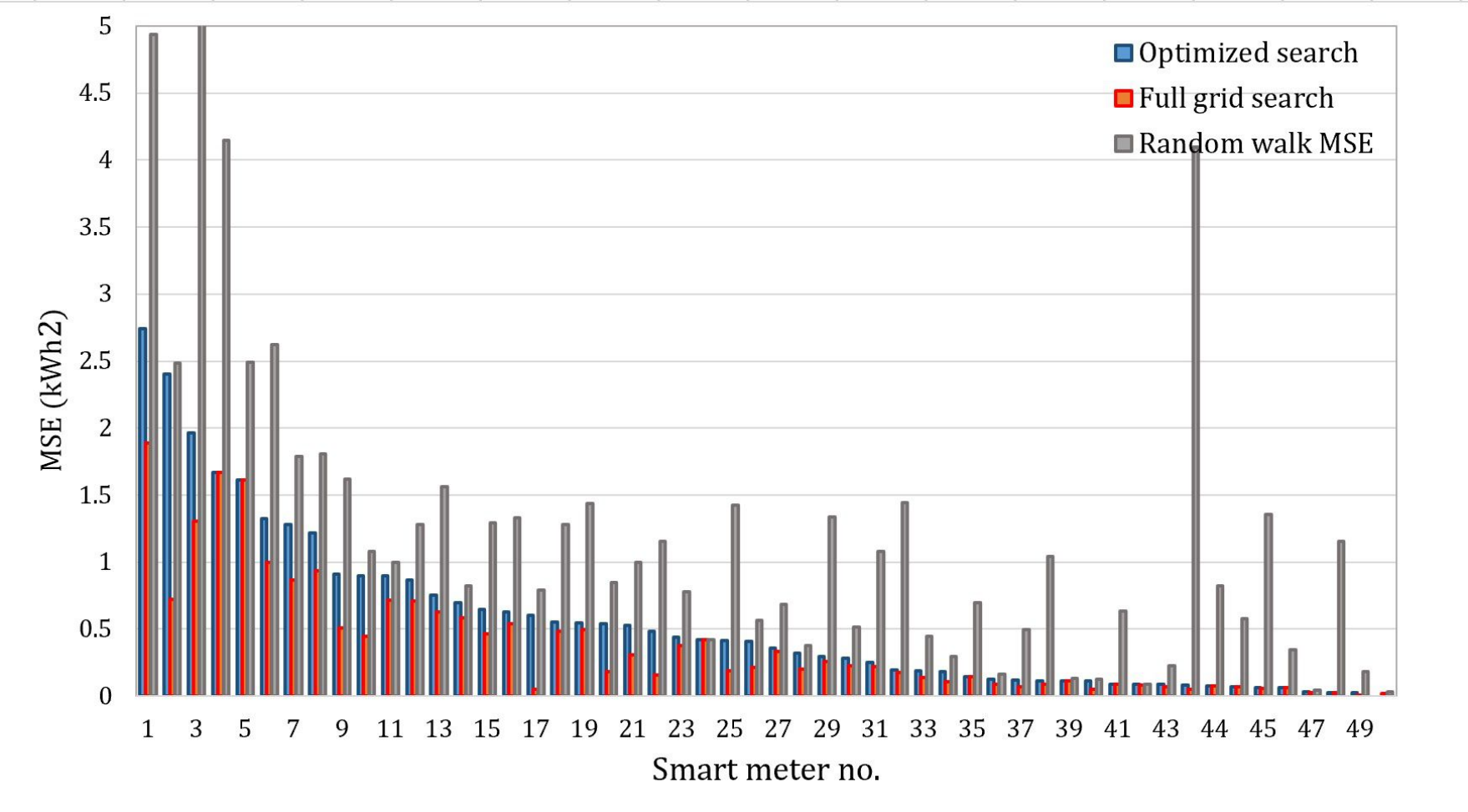}
\caption{MSE for 50 apartments - random walk, reduced and full grid search}
\label{fig:mse}
\end{figure}

The execution performance of the proposed ARIMA model order selection algorithm is presented in this section. Fig. \ref{fig:mse} shows the value of the MSE for every residential customer for full and reduced grid search, compared to random walk results. A random walk is defined as a process where the current value of a variable is composed of the past value plus an error term defined as a white noise (a normal variable with zero mean and variance one). The values are independent from one another and sorted in decreasing order of MSE for reduced grid search.

We can see that the reduced grid search approach has a higher error per customer than the full grid search, which is expected due to the reduced search space. However, we can also observe that the performance of the reduced grid search for the majority of the loads in terms of accuracy is close to the one achieved by the full grid search. In Table 1, the statistics of an increase in values for reduced grid search compared to the full grid search are presented. It can be seen that for the 50 apartments analyzed, the median of the error increase is at $22.54$\% for reduced grid search when compared to the full grid search approach.

\begin{table}
\caption{\label{tab:table-name}MSE statistics of reduced grid search compared to the full grid search (\%)}
\begin{center}
 \begin{tabular}{||c c c c c c ||} 
 \hline
 Min & 25\% & 50\% &75\% & Max & Average \\ [0.5ex] 
 \hline
 0 & 9.84 & 22.54 & 70.25 & 1100 & 82.15\\ [1ex] 
 \hline
\end{tabular}
\end{center}
\end{table}

\begin{figure}
\centering
\includegraphics [width=0.5\textwidth] {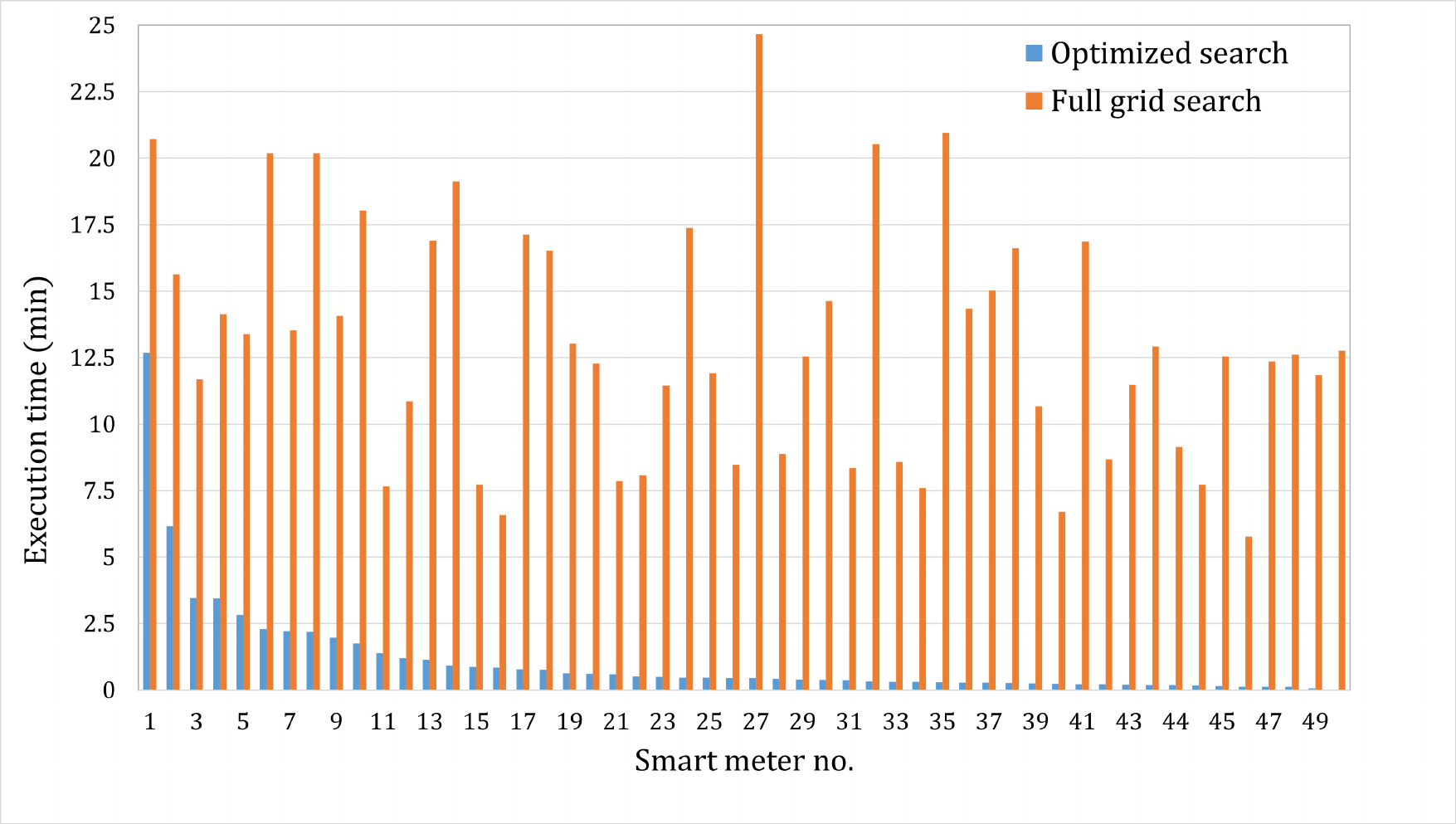}
\caption{Execution time comparison for reduced and full grid search}
\label{fig:ex_time}
\end{figure}

As the execution of this model order selection is expected to be carried out at the customer premises on low-level hardware, computation effort should be kept small. We see in Fig. \ref{fig:ex_time} that the execution time for the reduced grid search significantly outperforms the full grid search for all of the dwellings. Clearly, the reduced grid search is more suitable for local, real-time execution, while, at the same time, relative model forecast errors remain reasonable for practical settings.

\begin{figure}
\centering
\includegraphics [width=0.4\textwidth] {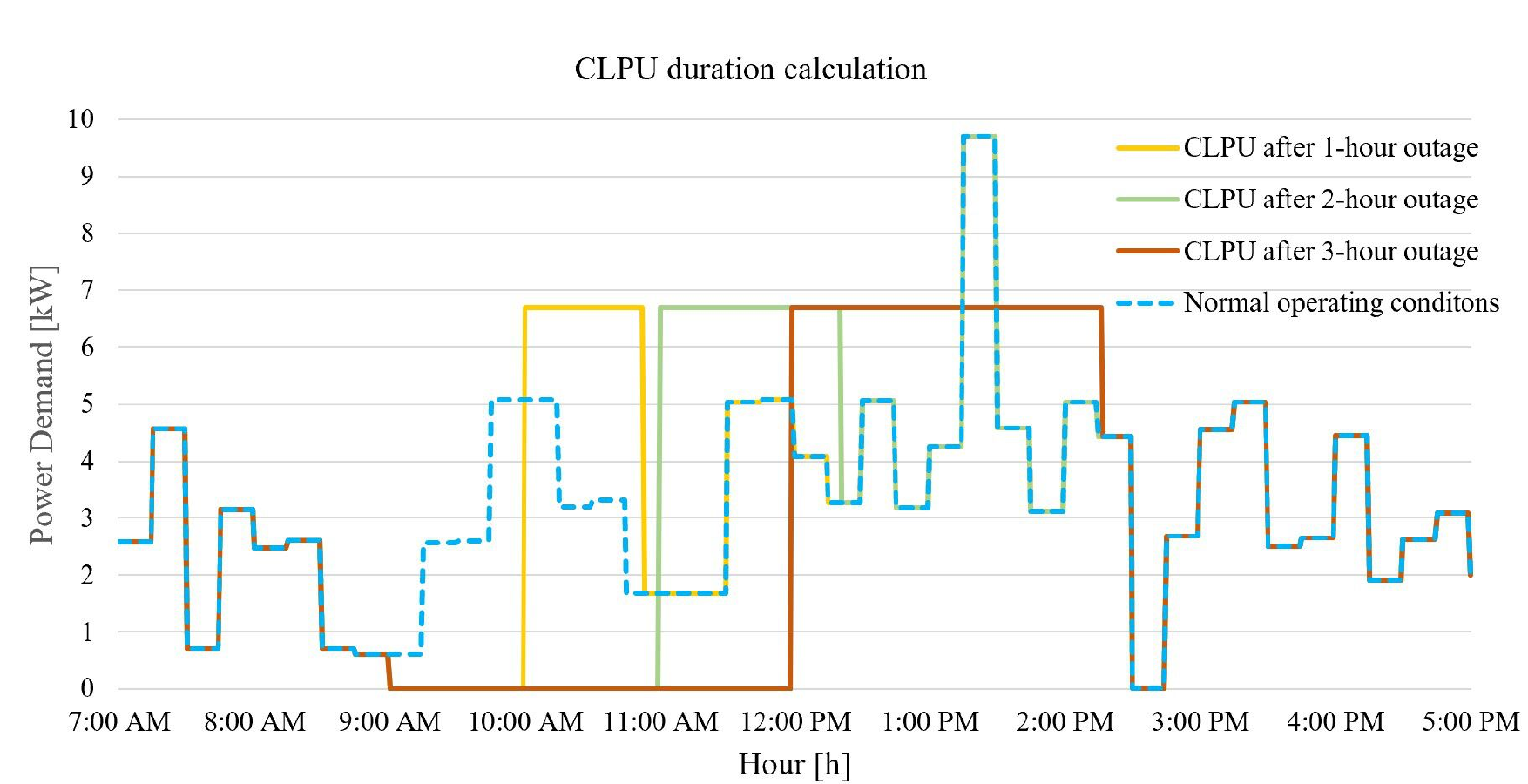}
\caption{Example of CLPU duration and peak forecasts for an outage starting at 9:00 AM}
\label{fig:clpudur}
\end{figure}

Fig. \ref{fig:clpudur} shows an example of calculation of the CLPU peak and duration using time-stamped load measurements for an outage starting at 9:00 AM lasting one, two and three hours and the demand in normal operating conditions. As expected, the forecasted CLPU energy is proportional to the outage duration. Moreover, the peak demands remain the same across outage durations since they are based on the peak CLPU forecated for outages starting at 9:00 AM. The difference between the CLPU peak and the normal operations demand seen between 1:00 and 2:00 PM is explained by the fact that the CLPU peak forecast is not tailored to capture future real-time operation power time series. We consider this not to be a major problem as in practice, ``normal'' demand patterns should be disrupted in the aftermath of outages lasting above one hour.

\subsection{Rolling Window Validation}
To verify the robustness of the developed approach when exposed to raw smart meter data, a rolling window forecast is executed for the one-year worth of data for the 50 apartments. By rolling window forecast, we consider the continuous execution of the proposed algorithm for a whole year, where model parameters are updated hourly using fresh data. With regularly updated model parameters, the CLPU peak and duration are predicted for outages of up to 12 hours and MSE is recorded. Considering that MSE increases with outage duration, the 12-hour MSE is taken as its worst-case. For each execution, one week of the most recent measurements is used. Considering the weekly periodicity of residential demand, the ARIMA model order (values of parameters $p$, $d$ and $q$) is revised on a weekly basis or in the case of the divergence of the MLE algorithm. 

\begin{figure}
\centering
\includegraphics [width=0.50\textwidth] {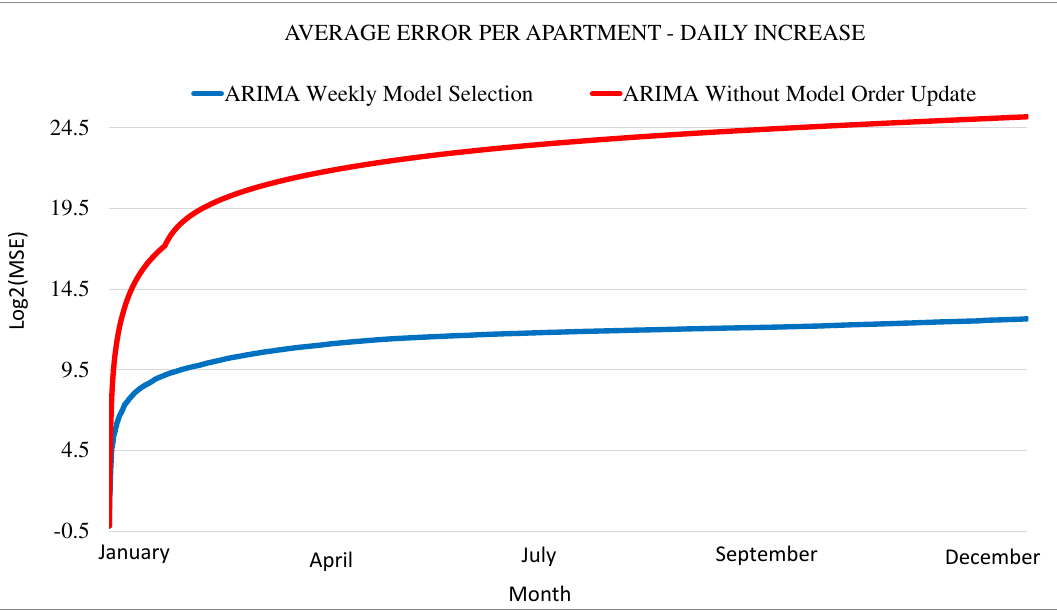}
\caption{Comparison of the ARIMA using weekly update of model order and ARIMA with no model update}
\label{fig:year}
\end{figure}

A weekly model update approach is compared to the one-year rolling window of ARIMA algorithm using the same, initially estimated, model order for the whole year. The average cumulative error comparison is shown in Fig \ref{fig:year}. We can observe that regular updates of the model order improve the accuracy significantly. The approach we propose can be executed independently for any unknown data set, without the need for user interference. This is one of the most significant properties of the local smart meter calculation unit.

\subsection{Performance Comparison Between ARIMA, LSTM and HWES}

The use of an ARIMA model can appear to be inadequate for performing this prediction task considering recent advances in deep learning techniques. In this section, we show the comparison between three common methods used for time-series forecasting for a 12-hour energy consumption modeling considering that only a limited amount of data is available. This assumption comes from the fact that smart meter devices that measure the consumption of residential customers have a limited set of functionalities, and therefore, limited memory.

\begin{figure}
\centering
\includegraphics [width=0.5\textwidth] {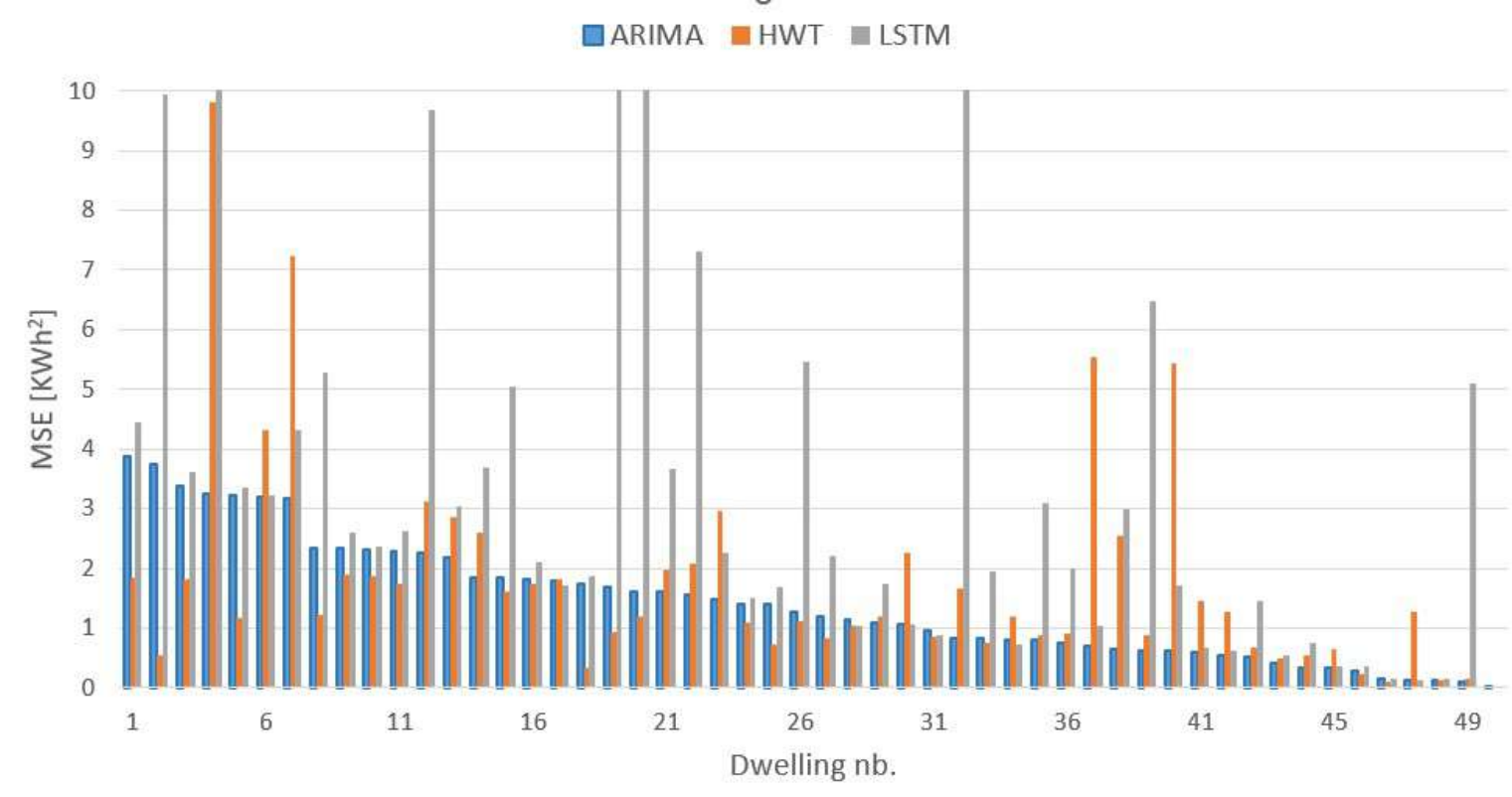}
\caption{Comparison of MSE for ARIMA, HWES and LSTM}
\label{fig:hwt}
\end{figure}

The accuracy and the execution speed of the proposed ARIMA approach have been compared with two common time series forecast approaches: long short term memory (LSTM) networks, described in Appendix D, and Holt-Winters exponential smoothing (HWES). In this case, the training and forecast are performed using only seven days of historical data, and forecast errors have been measured for the 50 randomly selected apartments. Fig.~\ref{fig:hwt} shows MSE value comparisons. It is important to note that because of the difference in the architecture between parametric and non-parametric approaches, out-of-the-box LSTM is not comparable to ARIMA or HWES. Therefore, the authors implemented a multistep LSTM based on the walking forward principle.

The performance of ARIMA has shown superior performance to LSTM in all cases mainly because neural network approaches are not performing well on small training sets. Exponential smoothing has an erratic performance, going from values that are much lower to much higher than ARIMA or even LSTM. Table~\ref{tab31} shows off MSE statistics for the three forecasting approaches. As observed from the Table 2, ARIMA presents the smallest standard deviation and the lowest error values on all percentiles. The median value is the same as for the HWES.
\begin{table}[!h]
\begin{center}
\caption{\label{tab31}MSE Statistics for the Three Forecasting Approaches}
\begin{tabular}{|c|c|c|c|}
\hline MSE Statistics & ARIMA & HWT & LSTM \\
\hline MIN & $0.01$ & $0.02$ & $0.01$ \\
\hline $10 \%$ & $0.28$ & $0.35$ & $0.39$ \\
\hline $50 \%$ & $1.20$ & $1.20$ & $2.20$ \\
\hline 90\% & $2.43$ & $3.23$ & $7.56$ \\
\hline MAX & $3.87$ & $9.80$ & $19.45$ \\
\hline STD & $1.03$ & $1.82$ & $3.85$ \\
\hline
\end{tabular}
\end{center}
\end{table}

\subsection{Performance Comparison with the Addition of Ambient Temperature Inputs}

\begin{figure}
\centering
\includegraphics [width=0.5\textwidth] {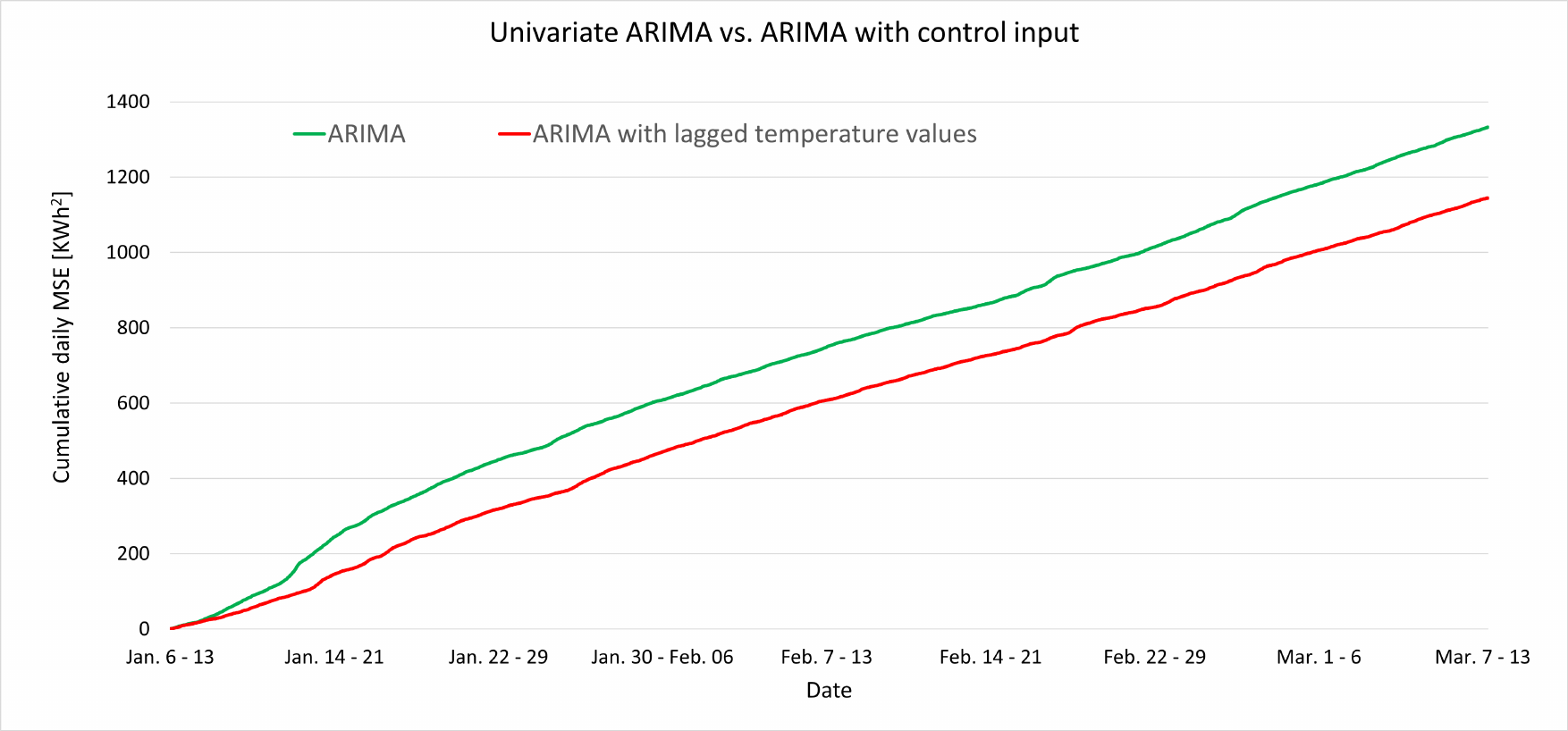}
\caption{Rolling window comparison of the ARIMA with and without lagged temperature values}
\label{fig:armax}
\end{figure}

CLPU performance is tightly related to thermostatic loading in a building. Therefore, lagged ambient temperatures could possibly help with the accuracy of CLPU forecasting. The accuracy of the ARIMA model with a temperature input using lagged temperature values as regressors, as seen in Appendix A, is compared against an univariate (demand only) ARIMA model. For this purpose, two months of rolling window validation is performed for 15 sets of smart meter data. The results are presented in Fig.~\ref{fig:armax} displaying daily cumulative errors.

The ARIMA model with temperature as a regressor has shown unsatisfactory performance in terms of convergence of the MLE algorithm and execution time. The utilization of temperature values has only slightly improved the accuracy of the forecast. However, the inclusion of $p$ lagged temperature values has significantly increased the complexity of the model and the procedure for ARIMA parameters selection would often diverge, even with an increased number of iterations. For example, two months of rolling window validation for ARIMA without control input executes in less than an hour on average, while the ARIMA model with lagged temperature values takes up to 12 hours. This is due to the fact that TCL energy consumption and lagged temperature values are co-linear, which essentially leads to the singularity of the normal equation used to determine the weights of the model.

\subsection{CLPU Forecasting Verification Using GridLab-D}
The proposed CLPU estimation approach of a single customer is verified using GridLab-D \cite{gridlabd}, a distribution network modelling software, which can implement the physical models of TCLs, as presented in Appendix \ref{appendix:gridlabd}. An outage range of one to 10 hours is simulated for the customer with electrical space and water heating using temperature data for the Boston, MA area in January. Energy consumption measurements are simulated at 15-minute intervals for one week. These are then used to train the ARIMA model. Energy not served during the outage is estimated using the ARIMA model. The peak power at which post-outage energy is consumed is calculated using auto-regression of the maximum power values for the previous seven days. The calculated peak value resulted in $16.51$ kW, while the average maximum power calculated by GridLab-D had a value $17.9$ kW.

\begin{figure}
\centering
\includegraphics [width=0.45\textwidth] {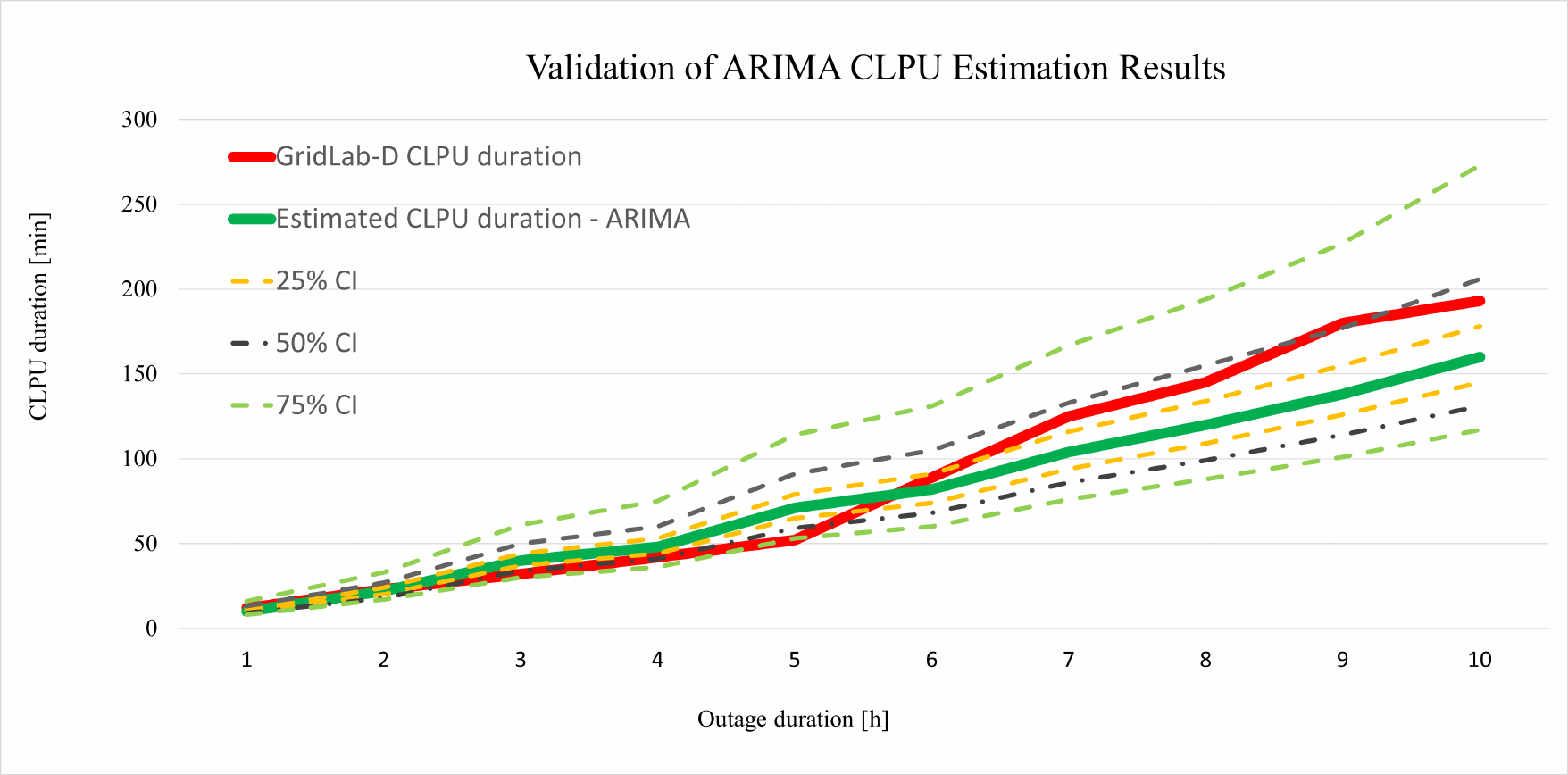}
\caption{Forecasted CLPU duration against GridLab-D for one dwelling}
\label{fig:gridlabd}
\end{figure}

Fig. \ref{fig:gridlabd} shows the CLPU duration calculated using the proposed CLPU estimation approach with $25\%$, $50\%$ and $75\%$ confidence interval (CI) of $P_{CLPU}(t)$ forecast, in comparison to the GridLab-D CLPU duration. We can observe that the GridLad-D calculations closely match the ARIMA results for short outage durations; moreover, later they all stay within its $50\%$ CI.

\section{Conclusion}
In this paper, we proposed an approach to estimate the local, customer-specific CLPU duration and peak using smart meter data. The duration of CLPU is forecasted based on the estimation of the energy not served during an outage. The energy not served is estimated using the ARIMA approach, where only timestamped past load measurements are used. The ARIMA model order is determined using a reduced grid search. We have shown that reduced grid search results in shorter execution time for model order selection while maintaining reasonable accuracy. Since the proposed framework is designated for local execution, low computational requirements are necessary; therefore, the approach relying on a reduced grid search yields acceptable results for reasonable computation efforts. We have observed that using temperature values significantly increases the execution time and model complexity while yielding minimal forecasting prediction improvements.

This work is foundational to performing adequate CLPU management during distribution system restoration in cold (and in warm) climates. With simple and low-computation customer-per-customer short-term energy forecasts, the proposed solution can be implemented along with single-customer service restoration, where the single-customer CLPU value plays a crucial role in restoration time and overload prevention.


\appendix
\section{CLPU Energy ARIMA Model with Lagged Ambient Temperatures}
\label{appendix:arimax}
The ARIMA energy consumption model can be extended with lagged temperature values to capture the temperature-dependence part of demand. The ARIMA model is extended with $p+1$ inputs, of which $p$ are lagged temperature values $u(t_0-p), \ldots, u(t_0-1)$ and the current temperature, $u(t_0)$
\begin{align}\label{ARIMAt:1}
    E_o(t_0,\tau_1) &= \sum_{j=0}^{r} \Big [ b(t_0+j) + E_o(t_0-d+j) \nonumber \\
    + & \sum_{i=1}^p \phi_i \tilde{E}_o(t_0-i+j) + \sum_{i=1}^q \theta_i b(t_0-i+j) \nonumber \\ 
    + & \sum_{i=1}^p \beta_i u(t_0-i+j) \Big ]
\end{align}
Obviously, if the temperature time series are non-stationary, they should be differenced accordingly. Therefore, the CLPU predictor in this case would be
\begin{align}\label{ARIMAt:2}
    \hat{E}_o(t_0,\tau_1) &= \sum_{j=0}^{r} \Big [ E_o(t_0-d+j) + \sum_{i=1}^p \phi_i \tilde{E}_o(t_0-i+j) \nonumber \\
    + & \sum_{i=1}^q \theta_i b(t_0-i+j) + \sum_{i=1}^p \beta_i u(t_0-i+j) \Big ]
\end{align}

\section{ACF and PACF Analysis} 
\label{appendix:acfpacf}
After a time series has been stationarized by differencing, the next step in fitting an ARIMA model is to determine how many AR ($p$) and MA ($q$) terms are needed to correct any autocorrelation that remains in the differenced series. By looking at the autocorrelation function (ACF) and partial autocorrelation function (PACF) plots of the differenced series, we can tentatively identify the numbers of AR and/or MA terms needed.

For an observed time series $\{x(1), x(2),\ldots, x(T)\}$, we denote the sample mean by $\bar{x}$. The sample lag-$h$ autocorrelation is given by:
\begin{equation}
\hat{\rho}_{h}=\frac{\sum_{t=h+1}^{T}\left(x(t)-\bar{x}\right)\left(x(t-h)-\bar{x}\right)}{\sum_{t=1}^{T}\left(x(t)-\bar{x}\right)^{2}}
\end{equation}
The sample lag-$h$ partial autocorrelation value is the estimated lag-$h$ coefficient in an AR model containing $h$ lags. The cut-off values, from which the limit for the order is set are calculated from confidence intervals for ACF and PACF, and they are determined from the minimum training data sample. The MA order is determined from ACF function confidence intervals ($CI$):

\begin{equation}
CI = \pm 1.96 \sqrt{\frac{1}{N}\left(1+2 \sum_{h=1}^{l} \hat{\rho}_{h}^{2}\right)}
\end{equation}
where the 1.96 factor is an approximate value of the quantile of the normal distribution for typical values of at 95\% confidence interval, $\hat{\rho}_{h}$ is the value of the sample ACF function for the $l^{\text{th}}$ lag, and  $N$ is a number of the measurements in the sample. On the other hand, the AR order is determined from the PACF function. Confidence intervals for PACF are calculated as
\begin{equation}
CI = \pm \frac{1.96}{\sqrt{N}}
\end{equation}

The values of ACF and PACF outside of the $CI$ are considered as significant ones and, based on the number of significant lags, the maximum values of the MA and AR orders are determined. The value of $CI$ influences the maximal order of parameters $p$ and $q$, and therefore, the ARIMA search space. The value of $CI = 95\%$ is selected as a good trade-off between accuracy, resulting in lower MSE with higher model orders, and on the other hand, lower search execution time which may result from limiting the model order.

\section{Long Short Term Memory (LSTM) Networks}
\begin{figure}
\centering
\includegraphics [width=0.40\textwidth] {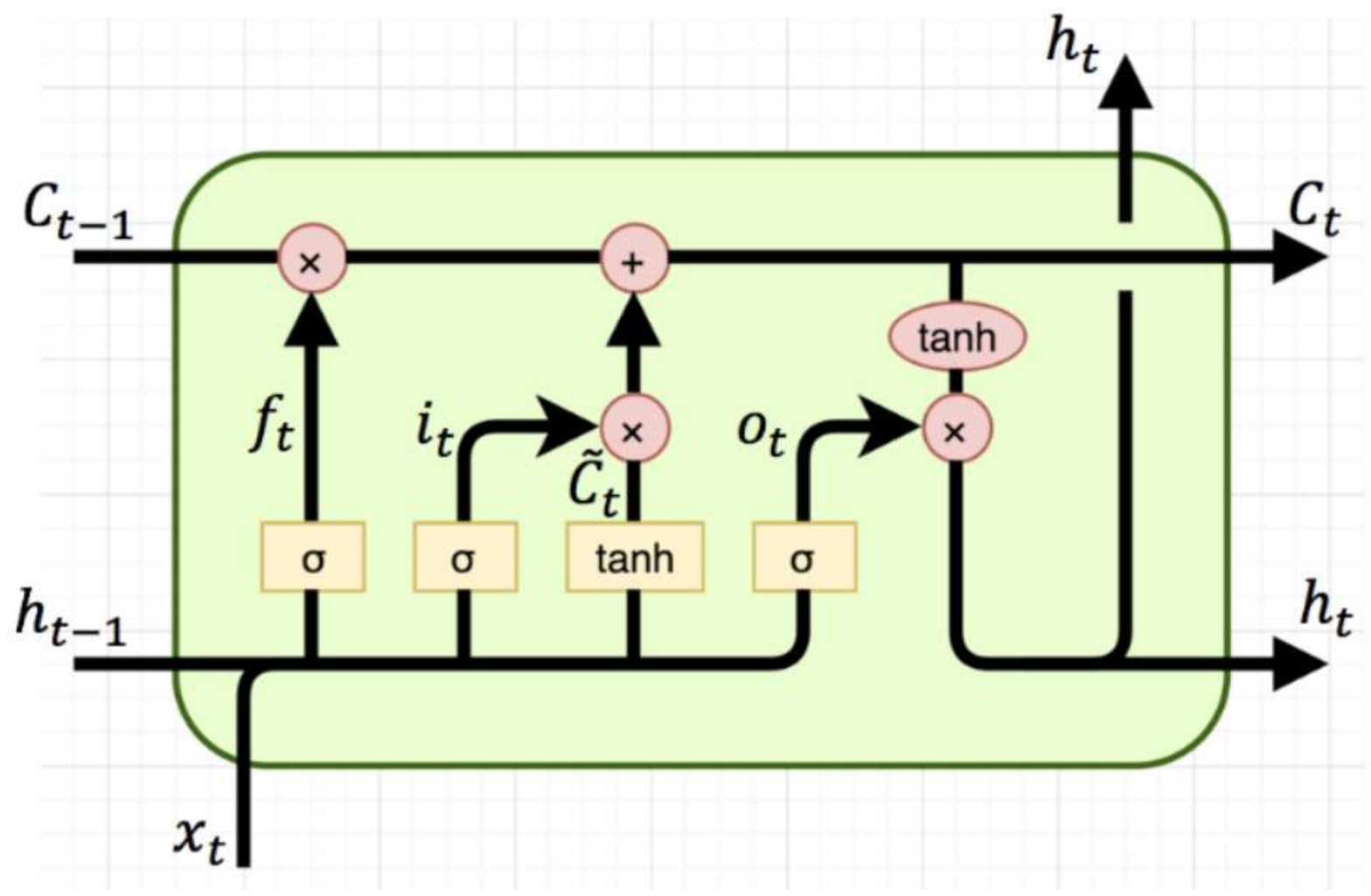}
\caption{Single cell of LSTM Network}
\label{fig:lstm}
\end{figure}

A LSTM is a special kind of RNN with additional features to memorize the sequence of data \cite{hochreiter1997long}.  
The representation of the LSTM having one memory block is depicted in Fig. \ref{fig:lstm} \cite{dey2021comparative}. It can be observed that the memory block contains input, output, and forget gates, which respectively represent write, read, and reset functions on each cell. Previous cell output ($h(t-1)$) and current input($x(t)$) are passed through the logistic(sigmoid) function. The long-term state ($c(t-1)$) is multiplied by the resulting value ($f(t)$). Forget gate decides how much of the previous data will be forgotten and how much of the previous data will be used in next steps. The sigmoid value obtained in the first step is also obtained here ($i(t)$) with the same inputs (Previous cell output ($h(t-1)$) and current input ($x(t)$). In addition, these inputs are also passed through the hyperbolic tangent (tanh) activation function and multiplied with the sigmoid output ($g(t)$). At this stage, (input gate) $i(t)$ mathematically expresses how important the  ($g(t)$) value is.
Mathematical expressions (Forget Gate and Input Gate) obtained in the first two cases are added and form the new cell state ($c(t)$). In addition, this result is regularized by passing through tanh and multiplied by the logistic(sigmoid) function result ($o(t)$) of the two inputs in the first two stages (Previous cell output ($h(t-1)$) and current input ($x(t)$). It returns the short-term state ($h(t)$). At this stage (output gate), the importance of long-term states is transferred to the system. 

\section{GridLab-D Load model}
\label{appendix:gridlabd}
The residential load in GridLab-D is modeled through the  ``house'' class which represents a thermodynamical load with integrated heating and cooling models. The load is modeled as a combination of ZIP loads, a heat pump and a water heater. In GridLab-D, the ``house'' thermal flow is calculated using the equivalent thermal parameter (ETP) model \cite{gridlab2013}. Given at the time $t$, the dynamics for the indoor air temperature $T_{A}$ and the building mass temperature $T_{M}$ are given
\begin{equation}
0=Q_{A}-U_{A}\left(T_{A}-T_{O}\right)-H_{M}\left(T_{A}-T_{M}\right)-C_{A} \frac{d}{d t} T_{A}
\end{equation}
\begin{equation}
    0=Q_{M}-H_{M}\left(T_{M}-T_{A}\right)-C_{M} \frac{d}{d t} T_{M}
\end{equation}
Here the constant values, that depend on the physical characteristics of the house and its environment, are: $T_O$ is the outdoor air temperature, $Q_A$ is the heat added to the indoor air, $Q_{M}$ is the heat added to the building mass, $U_{A}$ is the building envelope conductivity to the indoor air, $C_{A}$ is the heat capacity of the indoor air, 
$H_{M}$ is the building mass conductivity to the indoor air, and $C_{M}$ is the heat capacity of the building mass.

\section*{Acknowledgements}
The authors acknowledge the financial support received from the Natural Sciences and Engineering Research Council of Canada, Ottawa, ON, Canada (RGPIN-2018-05810; IRC Hydro-Québec) and from IVADO, Montreal, QC, Canada (BRDV-AUTR-201-2017-109).

\printcredits

\bibliography{cas-refs1}
\bibliographystyle{IEEEtran}




\bio{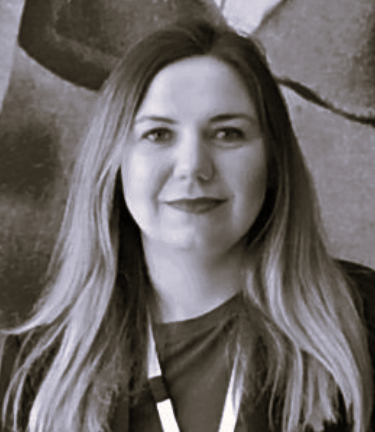}
Sanja Bajic received the B.Eng.(Hon.) and the M.Eng. in electrical engineering from University of Novi Sad, Serbia, in 2013 and 2014, respectively. She is currently pursuing her Ph.D. degree at McGill University, Montreal, Canada in Electrical Engineering and she is a member of Electric Energy Systems Laboratory. Her research interests include power system optimization, operation and planning and smart grids. Ms. Bajic is also a full-time researcher at the Hydro-Québec research institute (CRHQ).
\endbio
\newpage
\bio{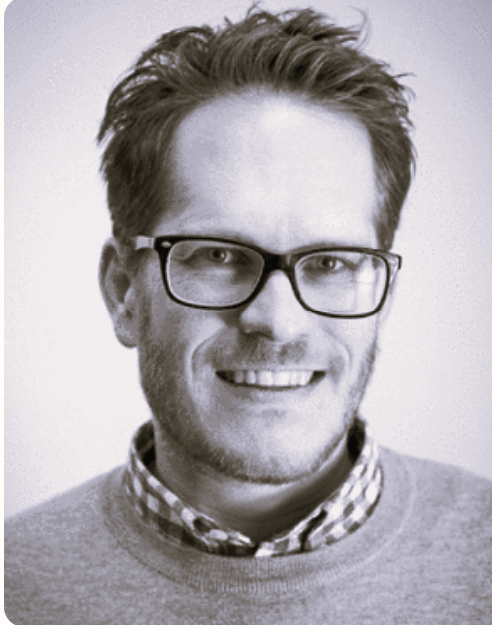}
François Bouffard received the B.Eng. (Hons.) and the Ph.D. degrees in electrical engineering from McGill University, Montreal, QC, Canada, in 2000 and 2006, respectively. From 2006 to 2010, he held a lectureship in electric power and energy with the School of Electrical and Electronic Engineering, The University of Manchester, Manchester, U.K. In 2010, he joined McGill University, where is currently an Associate Professor, William Dawson Scholar and Associate Chair of Undergraduate Affairs with the Department of Electrical and Computer Engineering. He is a licensed Engineer in the province of Québec, Canada. His research and teaching expertise include the fields of low-carbon power and energy system modeling, economics, reliability, control and optimization. He is a Member of the IEEE Power and Energy Society (PES). During 2009–2018, he was on the Editorial Board of IEEE Transactions on Power Systems. He is currently the Chair of the Power System Operations, Planning and Economics committee of IEEE PES.
\endbio

\bio{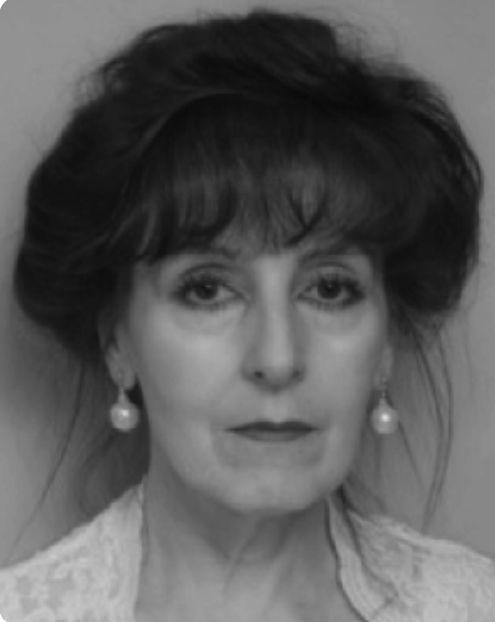}
Hannah Michalska received the M.Sc. degree in applied mathematics and the Ph.D. degree in control systems from the Imperial College of Science, Technology, and Medicine, London, U.K., in 1989. She was a Post-Doctoral Fellow with the Imperial College Research Centre on Process Systems. In 1992, she took the post of Assistant Professor and then an Associate Professor with the Department of Electrical and Computer Engineering, McGill University, Montreal. Her research interests include control of nonlinear systems, robotics, stabilization, and data fusion
\endbio

\bio{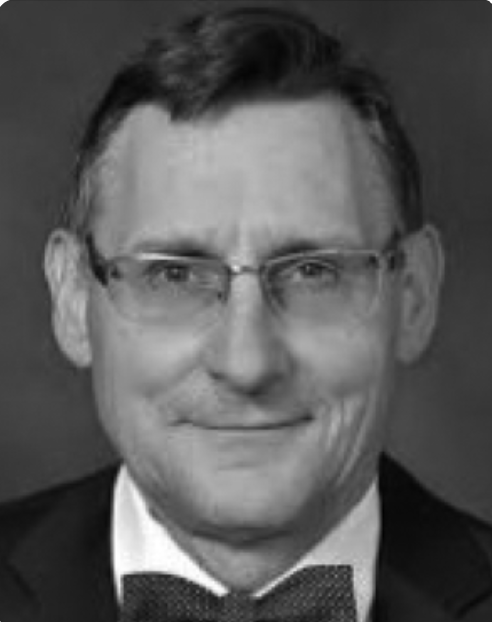}
Géza Joós received the M.Eng. and Ph.D. degrees in electrical engineering from McGill University, Montreal, QC, Canada. He has been a Professor with the Department of Electrical and Computer Engineering Department, McGill University, since 2001. He holds a Canada Research Chair in powering information technologies, since 2004. He was with ABB, the Université du Québec, Quebec, QC, Canada, and Concordia University, Montreal, QC, Canada. His research interests include distributed energy resources, including renewable energy resources, advanced distribution systems and microgrids. He is active in IEEE Power and Energy Society working groups on power electronic applications to power systems and IEEE Standards Association working groups on distributed energy resources. He is a Fellow of CIGRE, the Canadian Academy of Engineering and the Engineering Institute of Canada
\endbio

\end{document}